\documentclass[intlimits,twoside,a4paper]{article}

\usepackage{amsmath,amssymb}
\usepackage{graphicx}
\usepackage{epstopdf}

\usepackage{color}
\usepackage[T2A]{fontenc}
\usepackage[cp1251]{inputenc}
\usepackage[eqsecnum]{cmpj2}
\issue{2015}{18}{2}{23603}
\doinumber{10.5488/CMP.18.23603}

\title{A new kinetic model for precipitation from solid solutions}

\author{O. Borysenko}
\address{National Science Center ``Kharkiv Institute of Physics and Technology'', 1 Akademichna St., \\61108 Kharkiv, Ukraine}

\date{Received September 16, 2014, in final form March 2, 2015}%

\begin{document}

\maketitle

\begin{abstract}
This model considers reversible elementary acts of migration of point defects (interstitials and/or vacancies) across the precipitate-matrix interface and enables one to derive equations for the rates of emission and absorption of solute atoms at the interface. The model predicts much stronger heterophase fluctuations and higher nucleation rates than the classical nucleation theory does. However, asymptotically, for large precipitate sizes and long ageing times, both models give the same results, being in agreement with the Lifshitz-Slyozov-Wagner theory of coarsening.
\pacs  64.60.Q- % Nucleation
, 64.75.Nx % Phase separation and segregation in solid solutions
, 68.35.Dv % Composition, segregation; defects and impurities
, 81.30.Mh % Solid-phase precipitation
\end{abstract}

Quantitative studies of the kinetics of precipitation from solid
solutions originate from the pioneering work by Ostwald~\cite{Ostwald}.
The late stage of this process, called coarsening, is governed by the
theory elaborated by Lifshitz, Slyozov~\cite{LS} and Wagner~\cite{Wagner} (LSW).
At the same time, the general theory of precipitation is mainly based on
the classical nucleation theory (CNT), including its modifications and
extensions (see e.g.,~\cite{Slezov} and references therein), which is still debated in some of its aspects (see e.g.,~\cite{SSA} and references therein). In particular, one of
the open questions is a method of determination of the rates of
emission and absorption in the master kinetic equation~\cite{SSA}.
The present work is aimed at solving this problem.

This paper introduces a new phenomenological kinetic model for precipitation from solid solutions, being an
extension of the recently published model of homogeneous semicoherent
interphase boundary~\cite{Borysenko}. It considers the microscopic elementary acts of the reversible atomic rearrangements at the precipitate-matrix interface mediated by point defects (PDs). This approach enables one to determine
the rates of emission and absorption in the master kinetic equation [see equations~(\ref{W_e_small}) and (\ref{W_a_small}) below].

Consider an interphase boundary (Gibbs interface) between a precipitate consisting of one type of atoms
 and a solution of these atoms in a solid inert matrix. Let the interface
between the precipitate and the matrix be coherent, i.e., most of the
atomic planes be continuous across it. Since the bulk physical
properties of such a heterophase structure are discontinuous across the
interface, the number density (concentration) profiles of PDs are expected
to be discontinuous as well. The PDs can penetrate across the interface
via thermal activation or some other mechanism. Therefore, the  transfer of PDs across
the interface can be considered as a reversible surface chemical
reaction.

A solute atom in the interstitial position located at one side of
the interface can transfer to the other side of the interface and vice
versa. This process can be represented in the form of a reversible
chemical reaction:
\begin{eqnarray} \label{int}
\rm{i}_{\rm{s}}^{\rm{p}} \rightleftharpoons \rm{i}_{\rm{s}}^{\rm{m}},
\end{eqnarray}
where $\rm{i}_{\rm{s}}^{\varphi}$ denotes an interstitial solute atom in the precipitate
($\varphi = \rm{p}$) or in the matrix ($\varphi = \rm{m}$).
In this case, the rate of transitions, represented by equation~(\ref{int}), in each direction, is proportional to the concentration
$c_{\rm{i}_{\rm{s}}}^{\varphi}$ of the interstitials
$\rm{i}_{\rm{s}}^{\varphi}$ in
the corresponding phase. The normal component of the flux of
 solute atoms across the interface via the interstitial mechanism is
as follows (hereinafter the normal unit vector is supposed to be
directed from the precipitate into the matrix):
\begin{eqnarray} \label{int_flux}
j_{\rm{i}_{\rm{s}}}=\beta _{\rm{i}_{\rm{s}}}^{\rm{p}}c_{\rm{i}_{\rm{s}}}^{\rm{p}} - \beta _{\rm{i}_{\rm{s}}}^{\rm{m}}c_{\rm{i}_{\rm{s}}}^{\rm{m}}\,,
\end{eqnarray}
$\beta _{\rm{i}_{\rm{s}}}^{\varphi}$ being a phenomenological transition kinetic coefficient for the solute interstitials in the corresponding phase.

A solute atom located at a regular lattice site at one side of the interface can
transfer to the neighbouring vacant site at the other side of the interface
and vice versa. This process can be represented in the form of a
reversible chemical reaction:
\begin{eqnarray} \label{vac}
\rm{l}_{\rm{s}}^{\rm{p}} + \rm{v}^{\rm{m}} \rightleftharpoons \rm{l}_{\rm{s}}^{\rm{m}} + \rm{v}^{\rm{p}},
\end{eqnarray}
where $\rm{l}_{\rm{s}}^{\rm{p}}$ is a solute atom
at a regular
lattice site of the precipitate;
$\rm{v}^{\rm{m}}$ is a vacant regular
lattice site in the matrix;
$\rm{l}_{\rm{s}}^{\rm{m}}$ is a solute atom
at a regular
lattice site of the matrix;
$\rm{v}^{\rm{p}}$ is a vacant regular
lattice site in the precipitate.
Therefore, the rate of transitions, represented by equation~(\ref{vac}), in each direction, should be bilinear in
concentrations of the corresponding reagents. The normal component of
the flux of solute atoms across the interface via the vacancy mechanism
is as follows:
\begin{eqnarray} \label{vac_flux}
j_{\rm{v}_{\rm{s}}}= \beta_{\rm{v}_{\rm{s}}}^{\rm{m}}c_{\rm{l}_{\rm{s}}}^{\rm{p}}c_{\rm{v}}^{\rm{m}} - \beta_{\rm{v}_{\rm{s}}}^{\rm{p}}c_{\rm{l}_{\rm{s}}}^{\rm{m}}c_{\rm{v}}^{\rm{p}}\,.
\end{eqnarray}
Here, $c_{\rm{l}_{\rm{s}}}^{\rm{p}}$ is the
concentration of solute atoms that belong to the
regular lattice sites of the precipitate, $c_{\rm{v}}^{\rm{m}}$ is
the concentration of vacancies in the matrix, $c_{\rm{l}_{\rm{s}}}^{\rm{m}}$ is the
concentration of solute atoms at the regular lattice
sites of the matrix, $c_{\rm{v}}^{\rm{p}}$ is the
concentration of vacancies in the precipitate
 and $\beta _{\rm{v}_{\rm{s}}}^{\varphi}$ is a phenomenological transition kinetic coefficient for vacancies in the corresponding phase.

An interstitial atom located at one side of the interface can recombine
with a vacancy located at the other side:
\begin{eqnarray} \label{rec}
\rm{i}_{\rm{s}}^{\rm{p}} + \rm{v}^{\rm{m}} \rightarrow  \rm{l}_{\rm{s}}^{\rm{m}}; \qquad
\rm{i}_{\rm{s}}^{\rm{m}} + \rm{v}^{\rm{p}} \rightarrow  \rm{l}_{\rm{s}}^{\rm{p}}\,.
\end{eqnarray}
These are irreversible reactions because an energy threshold for
production of the Frenkel pairs is usually large. The normal component of
the flux of solute atoms via
the recombination mechanism (\ref{rec}) is as
follows:
\begin{eqnarray} \label{rec_flux}
j_{\rm{R}_{\rm{s}}}=\alpha_{\rm{R}_{\rm{s}}}^{\rm{p}}c_{\rm{i}_{\rm{s}}}^{\rm{p}}c_{\rm{v}}^{\rm{m}} - \alpha_{\rm{R}_{\rm{s}}}^{\rm{m}}c_{\rm{i}_{\rm{s}}}^{\rm{m}}c_{\rm{v}}^{\rm{p}}\,,
\end{eqnarray}
where $\alpha_{\rm{R}_{\rm{s}}}^{\varphi}$ is a phenomenological
recombination kinetic coefficient in the corresponding phase.

The total flux of solute atoms across the interface is
a sum of the contributions given by equations~(\ref{int_flux}), (\ref{vac_flux}) and (\ref{rec_flux}):
\begin{eqnarray} \label{tot_flux}
j_{\rm{s}}=j_{\rm{i}_{\rm{s}}}+j_{\rm{v}_{\rm{s}}}+j_{\rm{R}_{\rm{s}}}\,.
\end{eqnarray}
A total concentration of solute atoms in the corresponding
phase consists of the concentrations of the atoms in both the interstitial
and regular positions: $c_{\rm{s}}^{\varphi}=c_{\rm{i}_{\rm{s}}}^{\varphi}+c_{\rm{l}_{\rm{s}}}^{\varphi}$.
One can consider the following relations between the concentrations of
solute atoms in different lattice positions:
\begin{eqnarray} \label{conc}
c_{\rm{i}_{\rm{s}}}^{\varphi}=x_{\rm{s}}^{\varphi}c_{\rm{s}}^{\varphi};\qquad
c_{\rm{l}_{\rm{s}}}^{\varphi}=\left(1-x_{\rm{s}}^{\varphi} \right)c_{\rm{s}}^{\varphi}\,,
\end{eqnarray}
where $x_{\rm{s}}^{\varphi}$
is a dimensionless constant taking its value from the range
$0 \leqslant  x_{\rm{s}}^{\varphi} \leqslant  1$. The lower and upper limiting values
correspond to the cases when the solute atoms reside only in the regular
and interstitial lattice positions, respectively.
Then, taking into account equations~(\ref{int_flux}), (\ref{vac_flux}), (\ref{rec_flux}) and (\ref{conc}),
one can represent equation~(\ref{tot_flux}) as follows:
\begin{eqnarray} \label{flux_long}
j_{\rm{s}}=c_{\rm{s}}^{\rm{p}} \left[\beta_{\rm{i}_{\rm{s}}}^{\rm{p}}x_{\rm{s}}^{\rm{p}}+\beta_{\rm{v}_{\rm{s}}}^{\rm{m}}c_{\rm{v}}^{\rm{m}}\left(1-x_{\rm{s}}^{\rm{p}}\right)+\alpha_{\rm{R}_{\rm{s}}}^{\rm{p}}x_{\rm{s}}^{\rm{p}}c_{\rm{v}}^{\rm{m}}\right] %\nonumber \\
-c_{\rm{s}}^{\rm{m}} \left[\beta_{\rm{i}_{\rm{s}}}^{\rm{m}}x_{\rm{s}}^{\rm{m}}+\beta_{\rm{v}_{\rm{s}}}^{\rm{p}}c_{\rm{v}}^{\rm{p}}\left(1-x_{\rm{s}}^{\rm{m}}\right)+\alpha_{\rm{R}_{\rm{s}}}^{\rm{m}}x_{\rm{s}}^{\rm{m}}c_{\rm{v}}^{\rm{p}}\right].
\end{eqnarray}

The state of kinetic equilibrium at the interface is determined by the
condition of solute balance between the precipitate and the matrix:
\begin{eqnarray} \label{flux_eq}
j_{\rm{s}}^{\rm{eq}}=0.
\end{eqnarray}
Taking into account equation~(\ref{flux_long}), one can find from
equation~(\ref{flux_eq}) the relation between the equilibrium solute
and PDs concentrations at the interface:
\begin{eqnarray} \label{conc_eq}
\frac{c_{\rm{s}}^{\rm{m}~\rm{eq}}}{c_{\rm{s}}^{\rm{p}~\rm{eq}}}=\frac{\beta_{\rm{i}_{\rm{s}}}^{\rm{p}}x_{\rm{s}}^{\rm{p}~\rm{eq}}+ \left[\beta_{\rm{v}_{\rm{s}}}^{\rm{m}}\left(1-x_{\rm{s}}^{\rm{p}~\rm{eq}}\right)+ \alpha_{\rm{R}_{\rm{s}}}^{\rm{p}}x_{\rm{s}}^{\rm{p}~\rm{eq}}\right]c_{\rm{v}}^{\rm{m}~\rm{eq}}}{\beta_{\rm{i}_{\rm{s}}}^{\rm{m}}x_{\rm{s}}^{\rm{m}~\rm{eq}}+ \left[\beta_{\rm{v}_{\rm{s}}}^{\rm{p}}\left(1-x_{\rm{s}}^{\rm{m}~\rm{eq}}\right)+ \alpha_{\rm{R}_{\rm{s}}}^{\rm{m}}x_{\rm{s}}^{\rm{m}~\rm{eq}}\right]c_{\rm{v}}^{\rm{p}~\rm{eq}}}\,. %\nonumber \\
\end{eqnarray}

Provided that the thermal equilibrium between the precipitate and the matrix holds
(which is usually true for solids that exhibit high thermal conductivity), the
conditions of kinetic and thermodynamic equilibrium should be equivalent.
Therefore,
equation~(\ref{conc_eq}) is equivalent to the Gibbs-Thomson relation
for the equilibrium solute concentration near the
interface:
\begin{eqnarray} \label{conc_GT}
c_{\rm{s}}^{\rm{m}~\rm{eq}}\left(r_{\rm{p}}\right)=c_{\rm{s}}^{\rm{m}~\rm{eq}}\exp \left( a \big{/} r_{\rm{p}} \right),
\end{eqnarray}
where $c_{\rm{s}}^{\rm{m}~\rm{eq}}$ is a thermodynamic equilibrium
solubility, $r_{\rm{p}}$ is the radius of the precipitate and
\begin{eqnarray} \label{a}
a=2\gamma \omega _{0}\big{/}k_{\rm{B}}T
\end{eqnarray}
is the Gibbs-Thomson parameter with a dimension of length.
Here, $\gamma$ is the coefficient of the surface tension at the interface, $\omega_{0}$ is the
mean atomic volume, $k_{\rm{B}}$ is the Boltzmann's
constant and $T$ is temperature.

Now, we consider the problem of solute diffusion in the matrix near the
spherical precipitate of  radius $r_{\rm{p}}$.
A steady-state solute concentration profile in the matrix is subjected
to the following diffusion equation:
\begin{eqnarray} \label{diff_eq}
{\rm{div}} j_{\rm{s}}^{\rm{m}}=0;\qquad
j_{\rm{s}}^{\rm{m}}=-D_{\rm{s}}^{\rm{m}} \nabla c_{\rm{s}}^{\rm{m}}\,,
\end{eqnarray}
where $D_{\rm{s}}^{\rm{m}}$ is the solute
diffusion coefficient in the matrix.

%The initial condition for equation~(\ref{diff_eq}) can be set as follows:
%
%\begin{eqnarray} \label{init_cond}
%c_{\rm{s}}^{\rm{m}}\left( r, 0 \right)=\bar{c}_{\rm{s}}^{\rm{m}},
%\end{eqnarray}
%
%where $\bar{c}_{\rm{s}}^{\rm{m}}$
%is an average concentration of single solute atoms (momomers) in the matrix.

The normal component of the solute flux across the interface is given by
equation~(\ref{flux_long}). In the first order in a deviation of the
solute concentration at the interface from its kinetic equilibrium value
(\ref{conc_eq}), equation~(\ref{flux_long}) becomes
\begin{eqnarray} \label{flux_short}
j_{\rm{s}}^{\rm{m}}\left(r_{\rm{p}} \right)=D_{\rm{s}}^{\rm{m}} \left[c_{\rm{s}}^{\rm{m}~\rm{eq}}\left(r_{\rm{p}}\right)-c_{\rm{s}}^{\rm{m}}\left(r_{\rm{p}} \right)\right] \big{/}l,
\end{eqnarray}
where
\begin{eqnarray} \label{l}
l=D_{\rm{s}}^{\rm{m}}\Big{/}\left\{\beta_{\rm{i}_{\rm{s}}}^{\rm{m}}x_{\rm{s}}^{\rm{m}~\rm{eq}}+\left[\beta_{\rm{v}_{\rm{s}}}^{\rm{p}}\left(1-x_{\rm{s}}^{\rm{m}~\rm{eq}}\right)+\alpha_{\rm{R}_{\rm{s}}}^{\rm{m}}x_{\rm{s}}^{\rm{m}~\rm{eq}}\right]c_{\rm{v}}^{\rm{p}~\rm{eq}}\right\} %\nonumber \\
\end{eqnarray}
is the model parameter with a dimension of length. One can notice a similarity between equation~(\ref{flux_short}) and the Ohm's law.
By this analogy, the parameter $D_{\rm{s}}^{\rm{m}}/l$ can be considered as a ``conductivity'' of the interface, which comprises the
interstitial, vacancy and recombination mechanisms of mobility of solute atoms.

One can consider the second boundary condition as follows:
\begin{eqnarray} \label{conc_mean}
c_{\rm{s}}^{\rm{m}}\left( \infty \right)=\bar{c}_{\rm{s}}^{\rm{m}},
\end{eqnarray}
where $\bar{c}_{\rm{s}}^{\rm{m}}$ is an average concentration of single solute atoms (momomers) in the matrix.

The solution of the diffusion equation (\ref{diff_eq}) with the
boundary conditions, given by equations~(\ref{flux_short}) and
(\ref{conc_mean}), gives:
%
%\begin{align} \label{conc_r_p_t}
%c_{\rm{s}}^{\rm{m}}\left( r_{\rm{p}}, t \right)&= \bar{c}_{\rm{s}}^{\rm{m}}+\frac{r_{\rm{p}}\left[ c_{\rm{s}}^{\rm{m}~\rm{eq}}\left( r_{\rm{p}} \right)-\bar{c}_{\rm{s}}^{\rm{m}} \right]}{ r_{\rm{p}}+l} \nonumber \\
%&\times \left[1-\exp\left(\frac{t}{t_{\rm{sat}}}\right) \cdot {\rm{erfc}}\left(\sqrt{\frac{t}{t_{\rm{sat}}}}\right)\right],
%\end{align}
%
%where
%
%\begin{eqnarray} \label{t_sat}
%t_{\rm{sat}}=\left(r_{\rm{p}}^{-1}+l^{-1}\right)^{-2}\Big{/}D_{\rm{s}}^{\rm{m}}
%\end{eqnarray}
%
%is a characteristic time for the solute concentration at the interface to reach its steady-state value
%
\begin{eqnarray} \label{conc_r_p}
c_{\rm{s}}^{\rm{m}}\left( r_{\rm{p}} \right)=\bar{c}_{\rm{s}}^{\rm{m}}+r_{\rm{p}}\left[ c_{\rm{s}}^{\rm{m}~\rm{eq}}\left( r_{\rm{p}} \right)-\bar{c}_{\rm{s}}^{\rm{m}} \right] \big{/} \left(r_{\rm{p}}+l\right).
\end{eqnarray}
%

%From equation~(\ref{conc_r_p}) one can see that
%in a general case of a finite value of $l$, the steady-state value of the solute concentration at the interface is different from the
%equilibrium one and saturates to it for $r_{\rm{p}} \gg l$. This is a specific feature of the model, originating from the boundary
%condition equation~(\ref{flux_short}).

A total number $N=4 \pi r_{\rm{p}}^{3} \big{/} 3 \omega_{0}$ of atoms entering the precipitate is subject to
the following kinetic equation:
\begin{eqnarray} \label{dN_dt}
\rd N\big{/}\rd t=-4\pi r_{\rm{p}}^2 j_{\rm{s}} \left( r_{\rm{p}} \right) %\nonumber \\
=-4\pi r_{\rm{p}}^2 D_{\rm{s}}^{\rm{m}} \left[c_{\rm{s}}^{\rm{m}~\rm{eq}}\left(r_{\rm{p}}\right)-c_{\rm{s}}^{\rm{m}}\left(r_{\rm{p}}\right)\right] \big{/} l.
\end{eqnarray}
This equation can be reformulated in terms of the rates of emission and absorption of solute atoms
[i.e., the rates of direct and inverse reactions (\ref{int}), (\ref{vac}) and (\ref{rec})] at the interface in the
following way:
\begin{eqnarray} \label{dN_dt_2}
\rd N\big{/}\rd t=w_{\rm{a}} - w_{\rm{e}}\,,
\end{eqnarray}
where
\begin{eqnarray} \label{w_e}
w_{\rm{e}}=4\pi r_{\rm{p}}^2 D_{\rm{s}}^{\rm{m}}c_{\rm{s}}^{\rm{m}~\rm{eq}}\left(r_{\rm{p}}\right) \big{/} l;
\end{eqnarray}
\begin{eqnarray} \label{w_a}
w_{\rm{a}}=4\pi r_{\rm{p}}^2 D_{\rm{s}}^{\rm{m}}c_{\rm{s}}^{\rm{m}}\left(r_{\rm{p}}\right) \big{/} l
\end{eqnarray}
are respectively the rates of emission and absorption of solute atoms obtained by decomposition of the
right-hand side of equation~(\ref{dN_dt}) into the negative and positive parts.

Herein below, we study the case of homogeneous precipitation from a solid
solution within the framework of the Becker-D\"oring approach \cite{BD}.
In further considerations it is convenient to change to a dimensionless
time variable
\begin{eqnarray} \label{tau}
\tau=t \cdot 4\pi r_{0} D_{\rm{s}}^{\rm{m}} c_{\rm{s}}^{\rm{m}~\rm{eq}},
\end{eqnarray}
where $r_{0}=\sqrt[3]{3\omega_{0}/4\pi}$.

A distribution function
%(volume density)
$g \left( N, \tau \right)$ of precipitates in the dimension space %with respect to the total number $N$ of the monomers in the precipitate
is
subject to the next kinetic (master) equation \cite{BD}, valid for $N>1$:
\begin{eqnarray} \label{master_G}
\rd g \left( N, \tau \right) \big{/} \rd \tau = J_{N-1, N} - J_{N, N+1};
\end{eqnarray}
\begin{eqnarray} \label{J}
J_{N-1, N}= w_{\rm{a}}\left(N-1 \right) g \left( N-1, \tau \right) - w_{\rm{e}}\left(N \right) g \left( N, \tau \right).
\end{eqnarray}
%
%where $w_{\rm{e}}\left(N \right)$ and $w_{\rm{a}}\left(N \right)$ are respectively the rates of emission
%and absorption of solute monomers at the precipitate-matrix interface.

As a boundary condition, the following expression is used:
\begin{eqnarray} \label{g_N_max}
g \left( N_{\rm{max}}, \tau \right) =0.
\end{eqnarray}
Here, $N_{\rm{max}}$ stands for the number of atoms in the biggest precipitate under consideration.
It is assumed that for all $N \geqslant  N_{\rm{max}}$ the distribution function is zero. According to Lifshitz and Slyozov
\cite{LS}, at the late stage of the precipitation process, the value of $N_{\rm{max}}$ grows linearly with time.
The results presented herein below in figures~\ref{Fig:2} and \ref{Fig:3} are obtained with $N_{\rm{max}}=10^{12}$.

The system of equations~(\ref{master_G}) must be supplemented with
an additional equation for the value
\begin{eqnarray} \label{G1}
g \left( 1, \tau \right) = \bar{c}_{\rm{s}}^{\rm{m}} \left( \tau \right) \big{/} c_{\rm{s}}^{\rm{m}~\rm{eq}},
\end{eqnarray}
in order to satisfy the law of
conservation of the total amount of solute atoms $q$ [see equation~(\ref{q}) below]:
\begin{eqnarray} \label{master_G1}
\rd g \left( 1, \tau \right) \big{/} \rd \tau = -\sum_{N=2}^{N_{\rm{max}}} N \rd g \left( N, \tau \right) \big{/} \rd \tau.
\end{eqnarray}

From equations~(\ref{w_e}), (\ref{w_a}), taking into account equations~(\ref{conc_GT}), (\ref{conc_r_p}) and time renormalization (\ref{tau}), one finds:
\begin{eqnarray} \label{W_e_small}
w_{\rm{e}}\left(N \right)=\frac{\sqrt[3]{N^{2}}}{\lambda} \exp \left( \frac{\alpha}{\sqrt[3]{N}} \right);
\end{eqnarray}
\begin{eqnarray} \label{W_a_small}
w_{\rm{a}}\left(N \right)=\frac{\sqrt[3]{N^{2}}}{\lambda} \exp \left( \frac{\alpha}{\sqrt[3]{N}} \right) %\times \nonumber \\
\left\{ 1+\left[ g\left(1, \tau \right) \exp \left( -\frac{\alpha}{\sqrt[3]{N}} \right)-1\right]\frac{ \lambda}{\lambda+\sqrt[3]{N}} \right\},
\end{eqnarray}
where $\alpha = a\big{/}r_{0}$ and $\lambda = l\big{/}r_{0}$\,.

Equations (\ref{W_e_small}) and (\ref{W_a_small}) for the rates of emission and absorption at the interface are the key result of the present model.
They need to be compared to the corresponding equations derived in the CNT (see e.g., \cite{Slezov} and references therein), when the interface kinetics is taken into account.
In the present notations, the CNT expressions for the rates of emission and absorption are as follows:
\begin{eqnarray} \label{W_e_CNT}
w_{\rm{e}}^{\rm{CNT}}\left(N \right)=\frac{\sqrt[3]{N^{2}}}{\lambda + \sqrt[3]{N}} \exp \left( \frac{\alpha}{\sqrt[3]{N}} \right);
\end{eqnarray}
\begin{eqnarray} \label{W_a_CNT}
w_{\rm{a}}^{\rm{CNT}}\left(N \right)=\frac{\sqrt[3]{N^{2}}}{\lambda + \sqrt[3]{N}} g\left(1, \tau \right).
\end{eqnarray}
From equation~(\ref{W_a_CNT}) one can see that $w_{\rm{a}}^{\rm{CNT}}=0$ for $\bar{c}_{\rm{s}}^{\rm{m}}=0$, while, according to equation~(\ref{conc_r_p}), the steady-state concentration of solute atoms at the interface remains finite in this case: $c_{\rm{s}}^{\rm{m}}\left( r_{\rm{p}} \right)={}$\linebreak $c_{\rm{s}}^{\rm{m}~\rm{eq}}\left( r_{\rm{p}} \right) r_{\rm{p}} \big{/} \left(r_{\rm{p}}+l\right)$. It means that, once the solute atom has crossed the interface (via one elementary jump), within the framework of CNT it has no chance to jump back. On the contrary, the present theory considers reversible elementary acts at the interface [see equations~(\ref{int}) and (\ref{vac})]. On the other hand, from general speculations it follows that the rate of emission (``evaporation'') should be proportional to the area of the interface, i.e., $w_{\rm{e}} \propto \sqrt[3]{N^{2}}$. In the present model, this condition is satisfied for any $N$ [see equation~(\ref{W_e_small})], while in CNT it is satisfied only for $\sqrt[3]{N} \ll \lambda$ [see equation~(\ref{W_e_CNT})].Therefore, by neglecting reversible elementary acts at the interface, CNT underestimates the rates of emission and absorption of solute atoms.  At the same time, the value $\rd N\big{/}\rd \tau = w_{\rm{a}}-w_{\rm{e}}$, which is determined by the diffusion-controlled net solute flux in the matrix, is equal both in CNT and in this model. That is why both models give the same result in the asymptotic coarsening regime, but differ in the range of ultrafine precipitates (see figure~\ref{Fig:2} below). It should be noted that the results of this model are asymptotically equivalent to those of the CNT for $\lambda \gg \sqrt[3]{N}$ [cf. equations (\ref{W_e_small}), (\ref{W_e_CNT}) and (\ref{W_a_small}), (\ref{W_a_CNT})]. Great values of the parameter $\lambda$ correspond to the interface-limited precipitation regime, when the ``conductivity'' of the interface is small compared to the bulk one.

Under the condition of a detailed balance, when the flux of precipitates in
the dimension space (\ref{J}) turns to zero for any
$N$:
\begin{eqnarray} \label{J_0}
J_{N-1, N}=0,\quad \forall N
\end{eqnarray}
a stationary distribution function
$g_{0}\left(N\right)$ is given by the expression:
\begin{eqnarray}\label{g_0}
g_{0}\left(N\right)=
\begin{cases}
g_{0}\left(1\right),\quad N=1; \\
g_{0}\left(1\right) \prod \limits_{i=2}^{N} w_{\rm{a}}\left(i-1\right)\big{/}w_{\rm{e}}\left(i\right), \quad N>1.
\end{cases}
\end{eqnarray}

Provided that $\lim_{N \to \infty}g_{0}\left(N\right)=0$, the condition (\ref{J_0}) may be satisfied in the range
$0 \leqslant g_{0}\left(1\right) \leqslant 1$, which corresponds to the case of undersaturated and saturated solute concentrations
[see equation~(\ref{G1})].

A total concentration of solute atoms in the matrix (expressed in the units
of $c_{\rm{s}}^{\rm{m}~\rm{eq}}$) can be calculated as
follows:
\begin{eqnarray} \label{q}
q = \sum_{N=1}^{N_{\rm{max}}} N  g \left( N \right).
\end{eqnarray}

In the limiting case $g_{0}^{*}\left(1\right) =1$, corresponding to the saturated solute concentration, equation~(\ref{q})
with $g \left( N \right)=g_{0}^{*}\left(N\right)$ can be utilized to calculate the total solubility limit, taking into
account both the solute monomers and heterophase fluctuations (subcritical precipitates).

\begin{figure}[ht]
\centerline{\includegraphics[width=0.5\textwidth]{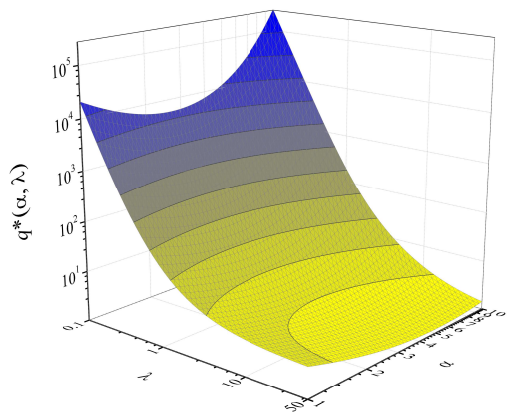}}
\caption{\label{Fig:1} (Color online) A total solubility limit $q^{*}$ (\ref{q*}) vs model parameters $\alpha$ and $\lambda$.}
\end{figure}
\begin{figure}[ht]
\centerline{\includegraphics[width=0.5\textwidth]{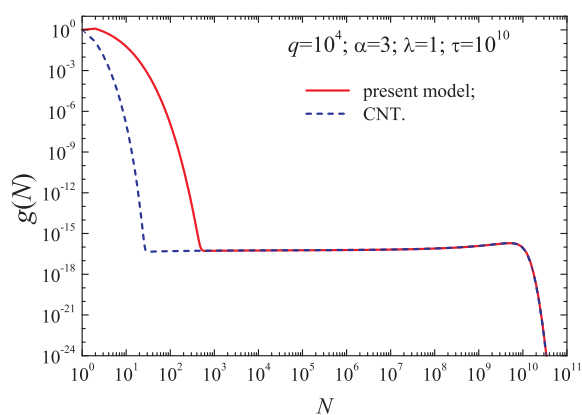}}
\caption{\label{Fig:2} (Color online) A distribution of precipitates at a given time calculated within the framework of CNT and this model.}
\end{figure}
\begin{figure}[!ht]
\centerline{\includegraphics[width=0.5\textwidth]{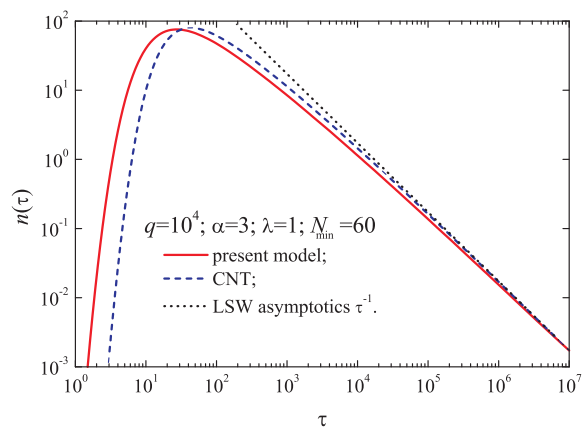}}
\caption{\label{Fig:3} (Color online) The concentration of precipitates as a function of time (\ref{n}) calculated within the framework of CNT and this model, together with the LSW asymptotic law $\tau^{-1}$.}
\end{figure}

Figure~\ref{Fig:1} shows the total solubility limit
\begin{eqnarray} \label{q*}
q^{*} = \sum_{N=1}^{N_{\rm{max}}} N  g_{0}^{*}\left(N\right)
\end{eqnarray}
as a function of two dimensionless model parameters $\alpha$ and $\lambda$ entering equations~(\ref{W_e_small}) and (\ref{W_a_small}). From figure~\ref{Fig:1} one can see that, within the present model, a
contribution from heterophase fluctuations to the total solubility limit, depending on the values of the model
parameters, may exceed the solubility of monomers  by several orders of magnitude.

Herein below, we compare the results of the present model for precipitation kinetics with those of CNT, for the same values of solute concentration in the matrix  $q=10^{4}$ and the model parameters $\alpha=3$ and $\lambda=1$.
In each calculation, the homogeneous state of a solid solution (only monomers, no precipitates) is taken as an initial condition.

Figure~\ref{Fig:2} shows a solution of the system of equations~(\ref{master_G}), (\ref{master_G1}) at $\tau=10^{10}$, with the rates of emission and absorption, given by this model [equations~(\ref{W_e_small}),  (\ref{W_a_small})] and CNT [equations~(\ref{W_e_CNT}), (\ref{W_a_CNT})]. The low-$N$ steep part of the curves describes heterophase fluctuations, while the
high-$N$ part describes the precipitates that evolve according to the LSW theory.
One can see that this model gives a much wider range of heterophase fluctuations than CNT does. At the same time, in the high-$N$
range, both models yield identical results. This result is in a qualitative agreement with several recent observations of subnanometer-sized clusters formed during ageing in supersaturated Fe-Cu \cite{Zhang}, \cite{Dmitrieva} and Ni-Al \cite{Singh} alloys.

Figure~\ref{Fig:3} shows the concentration of precipitates in the range $N_{\rm{min}} \leqslant  N \leqslant  N_{\rm{max}}$:
\begin{eqnarray} \label{n}
n \left( \tau \right) = \sum_{N=N_{\rm{min}}}^{N_{\rm{max}}}  g \left( N, \tau \right),
\end{eqnarray}
where $N_{\rm{min}}$ is a lower limit cutoff, practically set by the resolution limit of an observation instrument.
One can see that within the present model, the nucleation stage of the precipitation process occurs earlier than within CNT, and at the coarsening stage, the asymptotic LSW power law $n(\tau) \propto \tau^{-1}$ is achieved within the present model later than within CNT.

In summary, the present model, based on the consideration of reversible elementary acts of migration of point defects across the precipitate-matrix interface in a solid solution, allows for a direct derivation of the rates of emission and absorption of solute atoms at the interface and, therefore, makes it possible to study the kinetics of homogeneous precipitation from solid solutions. Compared with the classical nucleation theory, this model predicts much stronger heterophase fluctuations and higher nucleation rates. The results obtained apply to the kinetics of phase transformations in any other system where the boundary condition of the type of equation~(\ref{flux_short}) is applicable.

\section*{Acknowledgements}

The author is grateful to Dr.~A.~Turkin for sharing his numerical code and to Dr.~A.~Abyzov for reading and commenting the manuscript.

\newpage

%
%% If you have problems with typesetting in ukrainian uncomment lines below.
%
% \lastpage
 % \end{document}

\ukrainianpart

\title{Нова кінетична модель осадження із твердих розчинів}
\author{О.~Борисенко}
\address{Національний науковий центр ``Харківський фізико-технічний інститут'', вул. Академічна, 1, 61108, Харків, Україна}

\makeukrtitle

\begin{abstract}
\tolerance=3000%
В цій моделі проведено розгляд оборотних елементарних актів міграції точкових дефектів (міжвузельних атомів та/або вакансій)
через міжфазну границю виділення-матриця, що дає змогу отримати рівняння для швидкостей вивільнення та поглинання атомів домішки на границі.
У порівнянні з класичною теорією нуклеації, ця модель передбачає значно сильніші гетерофазні флуктуації та більші швидкості нуклеації. Проте асимптотично, для
великих розмірів виділень та довгого часу старіння, обидві моделі дають однакові результати, що збігаються з результатами теорії Ліфшиця-Сльозова-Вагнера.
\keywords гомогенна нуклеація, кінетика фазових перетворень, осадження, точкові дефекти, міжфазна границя

\end{abstract}


\begin{thebibliography}{99}

\bibitem{Ostwald} Ostwald~W., Z. Phys. Chem., 1897, \textbf{22}, 289.

\bibitem{LS} Lifshitz~I.M., Slyozov~V.V., J. Phys. Chem. Solids, 1961, \textbf{19}, 35; \bibdoi{10.1016/0022-3697(61)90054-3}.

\bibitem{Wagner} Wagner~C., Z. Electrochem., 1961, \textbf{65}, 581; \bibdoi{10.1002/bbpc.19610650704}.

\bibitem{Slezov} Slezov~V.V., Kinetics of First-Order Phase Transitions, WILEY-VCH Verlag GmbH \& Co. KGaA, Weinheim, 2009; \doi{10.1002/9783527627769}.

\bibitem{SSA} Schmelzer~J.W.P., Slezov~V.V., Abyzov~A.S., In: Nucleation Theory and Applications, Schmelzer~J.W.P. (Ed.), WILEY-VCH Verlag GmbH \& Co. KGaA, Weinheim, 2005, 39--73; \doi{10.1002/3527604790.ch3}.

\bibitem{Borysenko} Borisenko~A., J. Nucl. Mater., 2011, \textbf{410}, 69; \bibdoi{10.1016/j.jnucmat.2010.12.315}.

\bibitem{BD} Becker~R., D\"oring~W., Ann. Phys., 1935, \textbf{416}, 719; \bibdoi{10.1002/andp.19354160806}.

\bibitem{Zhang} Zhang~Z.W., Liu~C.T., Wang~X.-L., Littrell~K.C., Miller~M.K., An~K., Chin~B.A., Phys. Rev. B, 2011, \textbf{84}, 174114; \bibdoi{10.1103/PhysRevB.84.174114}.

\bibitem{Dmitrieva} Dmitrieva~O., Choi~P., Ponge~D., Raabe~D., Tillack~N., Hickel~T., Neugebauer~J., In: Scientific Report 2009/2010, Max-Planck-Institut fur Eisenforschung GmbH, Dusseldorf, 2010, 103.

\bibitem{Singh} Singh~A.R.P., Mechanisms of Ordered Gamma Prime Precipitation in Nickel Base Superalloys, Ph.D. thesis, University of North Texas, 2011.

\end{thebibliography}
\end{document}